\documentclass[%
superscriptaddress,
amsmath,amssymb,
aps,
prb,twocolumn,
]{revtex4-1}

\usepackage{graphicx}
\usepackage{dcolumn}
\usepackage{bm}

\usepackage{color}

\begin{document}

\preprint{APS/123-QED}

\title{
Incommensurate-Commensurate Magnetic Phase Transition in SmRu$_{2}$Al$_{10}$
}

\author{Shun Takai}
\affiliation{Department of Quantum Matter, AdSM, Hiroshima University, Higashi-Hiroshima 739-8530, Japan}
\author{Takeshi Matsumura}
\email[]{tmatsu@hiroshima-u.ac.jp}
\affiliation{Department of Quantum Matter, AdSM, Hiroshima University, Higashi-Hiroshima 739-8530, Japan}
\affiliation{Institute for Advanced Materials Research, Hiroshima University, Higashi-Hiroshima 739-8530, Japan}
\author{Hiroshi Tanida}
\affiliation{Department of Quantum Matter, AdSM, Hiroshima University, Higashi-Hiroshima 739-8530, Japan}
\author{Masafumi Sera}
\affiliation{Department of Quantum Matter, AdSM, Hiroshima University, Higashi-Hiroshima 739-8530, Japan}
\affiliation{Institute for Advanced Materials Research, Hiroshima University, Higashi-Hiroshima 739-8530, Japan}

\date{\today}

\begin{abstract}
Magnetic properties of single crystalline SmRu$_{2}$Al$_{10}$ have been investigated by electrical resistivity, magnetic susceptibility, and specific heat. 
We have confirmed the successive magnetic phase transitions at $T_{\text{N}}=12.3$ K and $T_{\text{M}}=5.6$ K. 
Resonant x-ray diffraction has also been performed to study the magnetic structures. 
Below $T_{\text{N}}$, the Sm$^{3+}$ moments order in an incommensurate structure with $\bm{q}_1=(0, 0.759, 0)$. 
The magnetic moments are oriented along the orthorhombic $b$ axis, which coincides with the magnetization easy axis in the paramagnetic phase. A very weak third harmonic peak is also observed at $\bm{q}_3=(0, 0.278, 0)$. 
The transition at $T_{\text{M}}$ is a lock-in transition to the commensurate structure described by $\bm{q}_1=(0, 0.75, 0)$. A well developed third harmonic peak is observed at $\bm{q}_3=(0, 0.25, 0)$. 
From the discussion of the magnetic structure, we propose that the long-range RKKY interaction plays an important role, in addition to the strong nearest neighbor antiferromagnetic interaction. 
\end{abstract}


\maketitle


\section{Introduction}
\label{sec:Intro}
Recently, a new type of Kondo semiconductor system, CeT$_2$Al$_{10}$ (T=Ru and Os) with orthorhombic YbFe$_2$Al$_{10}$-type structure (space group $Cmcm$, \#63),\cite{Thiede98} 
has attracted much interest because of its unconventional combination of strong $c$-$f$ hybridization and long-range magnetic order.~\cite{Muro09,Nishioka09,Strydom09,Tanida10a} 
The most prominent feature is its high transition temperature, $T_0$=27.3~K for CeRu$_2$Al$_{10}$ and 28.7~K for CeOs$_2$Al$_{10}$, respectively, which are much higher than $T_{\text{N}}$=16~K of GdRu$_2$Al$_{10}$.~\cite{Sera13Gd} 

There are several noteworthy properties associated with this phase transition. 
In CeRu$_2$Al$_{10}$, on one hand an antiferromagnetic order develops below $T_0$, where the magnetic moments of $0.34$ $\mu_{\text{B}}$ align along the $c$ axis with a propagation vector $\bm{q}=(0, 1, 0)$.~\cite{Khalyavin10} 
On the other hand, the magnetic susceptibility shows a spin-singlet like behavior below $T_0$; all the susceptibilities along the three crystallographic axes, $\chi_a$, $\chi_b$, and $\chi_c$, exhibit a steep decrease below $T_0$, in spite of the fact that $\chi_a$ and $\chi_b$ are perpendicular susceptibilities.~\cite{Tanida10b} 
Although the opening of a spin gap and charge gap in the ordered phase seems to be associated with this phase transition,~\cite{Robert10,Kimura11} the detailed mechanism has not yet been clarified. 
It is also anomalous that the ordered moment aligns along the $c$ axis regardless of the single ion anisotropy of  $\chi_a \gg \chi_c \gg \chi_b$ in the paramagnetic state.
Furthermore, the direction of the ordered moment can be tuned by introducing various kinds of perturbation such as magnetic field and atomic substitution.~\cite{Kondo11,Kondo13,Tanida13,Khalyavin13,Kobayashi14,Tanida14,Bhattacharyya14} 
It is currently interpreted that a strongly anisotropic $c$-$f$ hybridization plays an important role in determining the direction of the ordered moment.~\cite{Bhattacharyya14,Robert12,Sera13} 
The phase transition in CeRu$_2$Al$_{10}$ thus cannot be ascribed only to the normal Ruderman-Kittel-Kasuya-Yosida (RKKY) interaction. In this sense, the transition temperature has historically been written as $T_0$, rather than $T_{\text{N}}$.

In isostructural NdFe$_2$Al$_{10}$, the Nd moments order along the $a$ axis below $T_{\text{N}}$=3.9~K. 
The moment direction is consistent with the single ion anisotropy in the paramagnetic phase.~\cite{Kunimori12} 
The basic magnetic properties can be understood within the framework of mean-field model calculation including crystalline-electric-field (CEF) splitting and exchange interactions. 
In this sense, NdFe$_2$Al$_{10}$ can be regarded as a normal rare-earth compound, where the magnetic properties are dominated by the CEF states and RKKY exchange interactions. 
The magnetic structure found in NdFe$_2$Al$_{10}$, however, is a square-wave like structure described by double-$\bm{q}$ components, $\bm{q}_1=(0, 3/4, 0)$ and $\bm{q}_3=(0, 1/4, 0)=3\bm{q}_1 + \bm{\tau}_{0\bar{2}0}$.~\cite{Robert14} 
This is different from the simple antiferromagnetic arrangement of up and down moments in CeRu$_2$Al$_{10}$, and seems to reflect the real nature of the RKKY interaction in the RT$_2$Al$_{10}$ (R=rare earth) compounds. 
A similar magnetic structure is also reported in TbFe$_2$Al$_{10}$, where the $\bm{q}$-vector is $(0, 4/5, 0)$.~\cite{Reehuis00,Reehuis03} 

In the present work, we have studied the magnetic properties of isostructural SmRu$_2$Al$_{10}$. 
Basic properties of this compound using a polycrystalline sample has been reported recently.\cite{Peratheepan15}  
We report the results of electrical resistivity, magnetic susceptibility, and specific heat measurements using a single crystalline sample. 
It is shown that successive phase transitions take place at $T_{\text{N}}$=12.3~K and at $T_{\text{M}}$=5.6~K, which confirms the previous report. 
It is also shown that the moments align along the $b$ axis, which is consistent with the single ion anisotropy in the paramagnetic phase. 

In order to clarify the magnetic structure in the ordered phases, we have performed resonant x-ray diffraction (RXD) experiments. 
Since Sm is a strong neutron absorbing element, RXD is more suited in this study than neutron diffraction. 
High space resolution of synchrotron x-rays is also an advantage, which has been successfully utilized in this work in detecting the shift of the peak position at $T_{\text{M}}$. 
On the other hand, it is difficult to estimate the absolute value of the ordered moment by RXD. 
We show that the square-wave like magnetic structure in the low-temperature commensurate phase can be described by $\bm{q}_1=(0, 3/4, 0)$ and $\bm{q}_3=(0, 1/4, 0)=3\bm{q}_1 + \bm{\tau}_{0\bar{2}0}$ in the same way as in NdFe$_2$Al$_{10}$. 
The incommensurate magnetic structure above $T_{\text{M}}$, which is characterized by a small shift of the peak position and a significant decrease of the third harmonic peak intensity, is also described. 

\section{Experiment}
\label{sec:Exp}
Single crystals of SmRu$_{2}$Al$_{10}$ were prepared by an Al flux method. 
Typical size of the obtained crystals was approximately $1\times 2\times 2$ mm$^3$. 
The result of x-ray powder diffraction was consistent with the lattice parameter in the literature: $a=9.1087$ \AA, $b=10.2456$ \AA, $a=9.1577$ \AA.\cite{Sera13}  
Specific heat and magnetic susceptibility were measured by PPMS and MPMS (Quantum Design), respectively. 
Electrical resistivity was measured by normal four probe ac method up to 14.5 T. 
Resonant x-ray diffraction experiment was performed at BL-3A of the Photon Factory, High Energy Accelerator Research Organization (KEK), Japan. 
The sample was attached to a cryostat so that the $0kl$ reciprocal lattice plane spans the scattering plane. 
We used x-ray energies near the $L_3$ absorption edge of Sm. Polarization analysis was performed using a Cu-220 reflection. 

\section{Results}
\label{sec:Results}
\subsection{Basic properties}
\begin{figure}[t]
\begin{center}
\includegraphics[width=7cm]{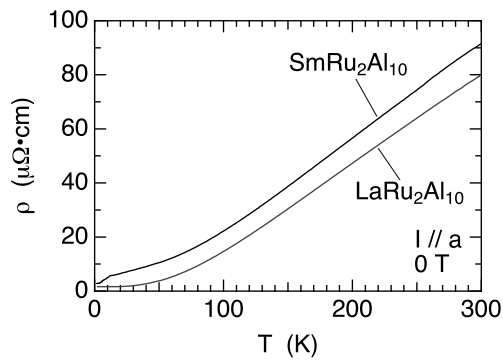}
\end{center}
\caption{Temperature dependence of the resistivity of SmRu$_{2}$Al$_{10}$ and LaRu$_{2}$Al$_{10}$ at zero field. }
\label{fig:fig1}
\end{figure}

\begin{figure}[t]
\begin{center}
\includegraphics[width=8.5cm]{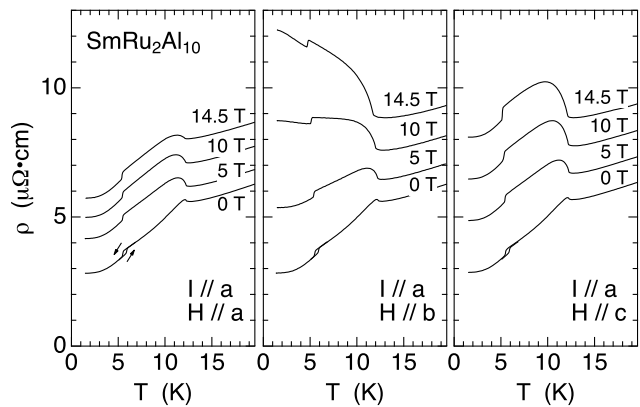}
\end{center}
\caption{Temperature dependences of the resistivity at several magnetic fields applied along the three principal directions. }
\label{fig:fig2}
\end{figure}

Figure \ref{fig:fig1} shows the temperature dependence of the resistivity ($\rho$) at zero magnetic field; $\rho(T)$ of LaRu$_2$Al$_{10}$ is also shown for reference. 
The $\rho(T)$ curve of SmRu$_2$Al$_{10}$ exhibits a smooth increase up to 300~K, which is almost parallel to the $\rho(T)$ curve of LaRu$_{2}$Al$_{10}$. 
This shows that the Kondo effect, and therefore the $c$-$f$ hybridization effect, is small in SmRu$_2$Al$_{10}$. 
Figure \ref{fig:fig2} shows $\rho(T)$ at several magnetic fields applied along $a$, $b$, and $c$ axes. 
At zero field, $\rho(T)$ exhibits a weak increase and kink at $T_{\text{N}}=12.3$~K, corresponding to an antiferromagnetic phase transition. 
At $T_{\text{M}}=5.6$~K, we observe another anomaly, where the resistivity slightly decreases with a jump and hysteresis, suggesting a first order transition. 

In applied magnetic fields, the resistivity generally shows a positive magnetoresistance. 
With increasing the field, the increase in resistivity at $T_{\text{N}}$ is more enhanced, particularly for $H\parallel b$ and $\parallel c$, whereas the anomaly at $T_{\text{M}}$ changes little. 
In addition, at 14.5T, $\rho$ continues to increase with decreasing temperature only for $H\parallel b$. 
This is probably due to an anisotropic super-zone gap appearing below $T_{\text{N}}$, which is related to the magnetic structure. Although this could provide valuable information on the Fermi surface structure, further study is necessary to extract an interpretation from this result, such as measuring $\rho(T)$ for the three field directions and the three current directions as performed in Ref.~\onlinecite{Yoshida15}. 
Finally, we note that the transition temperatures are hardly affected by the field. 
This is probably because the $g$-factor of Sm$^{3+}$ is small ($g$=2/7). 
This point will be discussed in Sec. \ref{subsec:DiscD}.

\begin{figure}[t]
\begin{center}
\includegraphics[width=8.5cm]{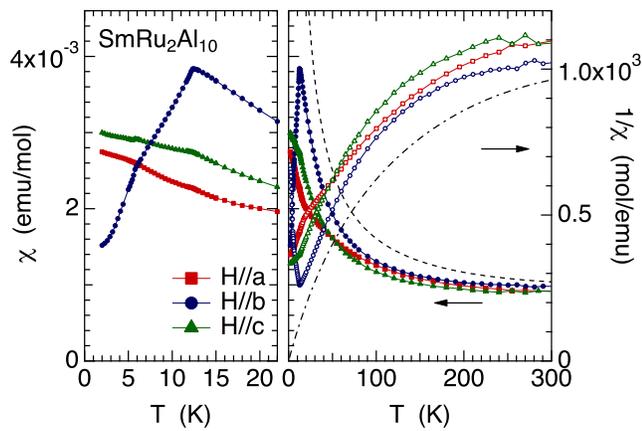}
\end{center}
\caption{(Color online) Temperature dependence of magnetic susceptibility for the three principal field directions measured at 1 T. 
Inverse magnetic susceptibility is also shown. 
The dashed and dot-dashed curves are the calculations of $\chi$ and $1/\chi$, respectively, for a free Sm$^{3+}$ ion with $J=5/2$ ground state and $J=7/2$ excited state at 1500~K. }
\label{fig:fig3}
\end{figure}

Figure \ref{fig:fig3} shows the temperature dependences of magnetic susceptibility $\chi_a$, $\chi_b$, and $\chi_c$ for $H \parallel a$, $b$, and $c$ axis, respectively. 
We see that $\chi_b$ is the largest in the paramagnetic phase and exhibits a clear cusp anomaly at $T_{\text{N}}$, whereas $\chi_a$ and $\chi_c$ are smaller than $\chi_b$ and continue to increase with decreasing temperature even below $T_{\text{N}}$. 
These results show that the $b$ axis is the easy axis of magnetization and the magnetic moments in the antiferromagnetic ordered phase are oriented along the $b$ axis ($\bm{\mu}_{\text{AF}} \parallel b$). 
A weak kink is also observed at $T_{\text{M}}$. 

The magnetic anisotropy of SmRu$_2$Al$_{10}$, $\chi_b > \chi_c > \chi_a$ and $\bm{\mu}_{\text{AF}} \parallel b$, is opposite to those of Rh-doped CeRu$_2$Al$_{10}$ and of NdRu$_2$Al$_{10}$, where $\chi_a > \chi_c > \chi_b$ and $\bm{\mu}_{\text{AF}} \parallel a$.~\cite{Kondo13,Tanida11} 
This is probably due to the difference in the sign of the second order Stevens factor. 
It is positive for Sm$^{3+}$, whereas it is negative for Ce$^{3+}$ and Nd$^{3+}$. 
This means that the $4f$ charge distribution of a Sm$^{3+}$ ion prefers to be elongated parallel to the magnetic moment. 
Since the charge distribution is coupled with the lattice, the magnetic anisotropy of SmRu$_2$Al$_{10}$ is opposite to those of Ce and Nd counterparts.  

As in most of the Sm based compounds, $1/\chi$ deviates from the Curie-Weiss $T$-linear behavior above $\sim 100$~K. 
This is due to the Van Vleck contribution from the $J=7/2$ excited multiplet. 
The calculated magnetic susceptibility of a free Sm$^{3+}$ ion, by assuming the $J=7/2$ energy level at 1500~K,\cite{Alekseev97} is shown by the dashed and dot-dashed curves in Fig.~\ref{fig:fig3}.
The calculated curve well explains the experimental data if we take into account an antiferromagnetic exchange interaction, which vertically shifts the calculated $1/\chi$ curve.

\begin{figure}[t]
\begin{center}
\includegraphics[width=7.5cm]{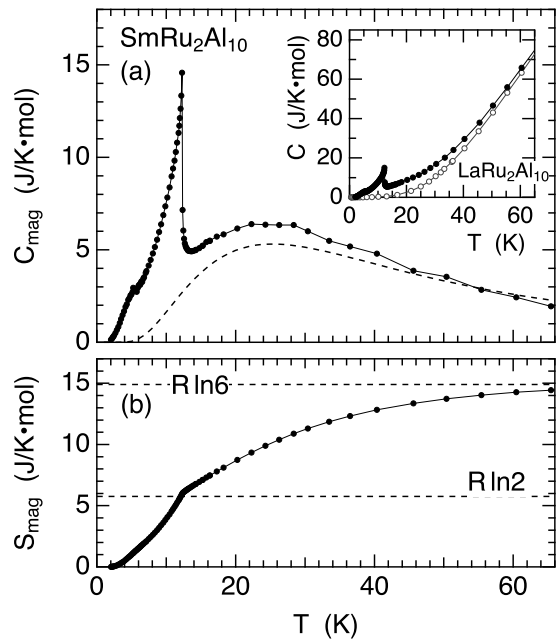}
\end{center}
\caption{(a) Temperature dependence of the magnetic specific heat obtained after subtracting the specific heat of LaRu$_2$Al$_{10}$. Inset shows the raw data of specific heat of SmRu$_2$Al$_{10}$ and LaRu$_2$Al$_{10}$. 
The dashed line shows a calculation of the Schottky-type specific heat obtained by assuming a CEF Hamiltonian with a level scheme of 0~K -- 47~K -- 95~K. 
(b) Temperature dependence of the magnetic entropy. }
\label{fig:fig4}
\end{figure}

Figure \ref{fig:fig4} shows the temperature dependence of magnetic specific heat and entropy at zero field, which almost reproduces the data reported in Ref.~\onlinecite{Peratheepan15}. 
The large $\lambda$-type anomaly in $C_{\text{mag}}(T)$ at $T_{\text{N}}$ reflects the magnetic ordering. 
The weak anomaly at $T_{\text{M}}$ shows that the entropy change at $T_{\text{M}}$ is small. 
In the paramagnetic state, a Schottky-type broad peak is observed in $C_{\text{mag}}(T)$ around 20~K. 
This anomaly can be ascribed to three Kramers doublets split from the $J=5/2$ multiplet. 
The calculated specific heat for a CEF level scheme of 0~K -- 47~K -- 95~K is represented by the dashed curve in Fig.~\ref{fig:fig4}(a). This level scheme will be discussed in Sec. \ref{subsec:DiscD} to explain the anisotropy of magnetic susceptibility. 
The magnetic entropy released at $T_{\text{N}}$ is $R \ln 2$, indicating that the CEF ground state is a well isolated doublet. 
At high temperatures, the entropy approaches $R \ln 6$. These are consistent with the above CEF level scheme.

\subsection{Resonant x-ray diffraction}
\begin{figure}[t]
\begin{center}
\includegraphics[width=8cm]{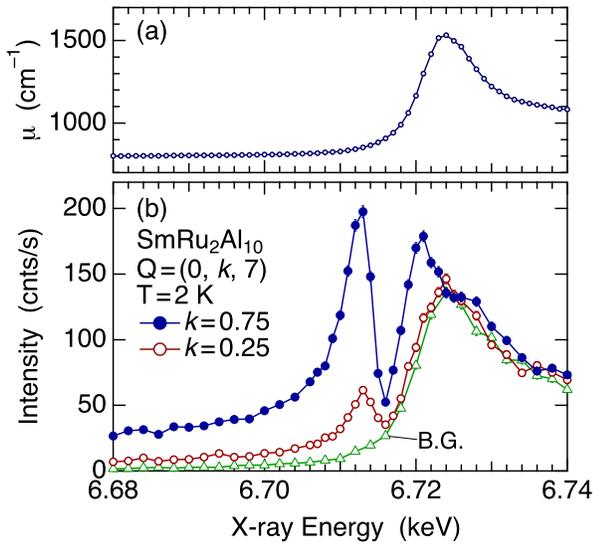}
\end{center}
\caption{(Color online) (a) Absorption coefficient obtained from the fluorescence spectrum. 
(b) X-ray energy dependences of the magnetic Bragg peak at $\bm{Q}=(0, 0.75, 7)$ and $(0, 0.25, 7)$ at 2~K without polarization analysis. 
The triangles represent the background. 
}
\label{fig:fig5}
\end{figure}

In the reciprocal lattice scan at the lowest temperature of 2~K, we have found clear diffraction peaks at positions represented by $\bm{Q}=\bm{q}+\bm{\tau}$, where $\bm{q}=(0, 0.75, 0)$ and $\bm{\tau}$ the reciprocal lattice vector of the fundamental lattice. 
We have also found peaks at the third harmonic positions corresponding to $3\bm{q}$, although the intensities were weaker than those of the first harmonic peaks. 
The energy dependences of the superlattice peaks at $\bm{Q}=(0, 0.75, 7)=\bm{q}+\bm{\tau}_{007}$ and at $\bm{Q}=(0, 0.25, 7)=3\bm{q}+\bm{\tau}_{0\bar{2}7}$ are shown in Fig.~\ref{fig:fig5}.
The intensity exhibits a resonant enhancement at 6.712 keV and 6.721 keV, corresponding to the $E2$ ($2p\leftrightarrow 4f$) and $E1$ ($2p\leftrightarrow 5d$) resonances, respectively.
The strong enhancement of the $E2$ resonance of magnetic dipole origin is also observed in SmB$_2$C$_2$ and SmRu$_4$P$_{12}$.\cite{Inami07,Matsumura14} 
Another broad peak around 6.728 keV, which is more clearly recognized in the first harmonic signal with stronger intensity, also belongs to the $E1$ resonance, and probably reflects the density of states of the $5d$ band. 
It is also noted that the energy-independent intensity below 6.70 keV is due to the nonresonant magnetic scattering.

\begin{figure}[t]
\begin{center}
\includegraphics[width=8cm]{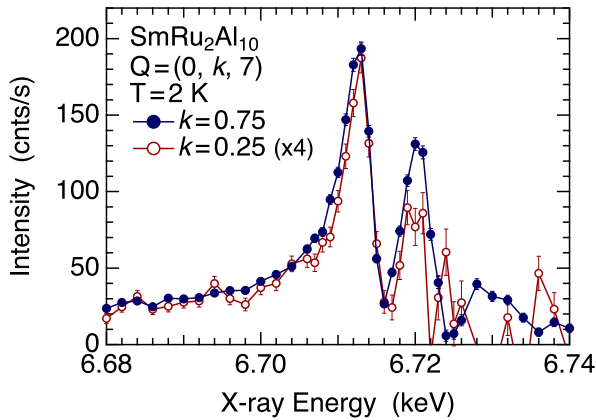}
\end{center} 
\caption{(Color online) X-ray energy dependences of the magnetic Bragg peak at $\bm{Q}=(0, 0.75, 7)$ and $(0, 0.25, 7)$ at 2~K without polarization analysis 
after subtracting the background and being corrected for absorption. The data for $k=0.25$ are multiplied by four. 
The error bars represent only the statistical error.}
\label{fig:fig6}
\end{figure}

In Fig.~\ref{fig:fig6}, we show the results of Fig.~\ref{fig:fig5}(b) after subtracting the background and being corrected for absorption, 
where the intensity is normalized at 6.68 keV. 
This figure shows that the intensity of the $3\bm{q}$ peak is $\sim0.25$ times as that of the first harmonic peak over the whole energy range from 6.68 keV to 6.72 keV. 
This intensity ratio will be discussed in Sec.~\ref{subsec:DiscA} in relation to a magnetic structure model.

\begin{figure}[t]
\begin{center}
\includegraphics[width=8.5cm]{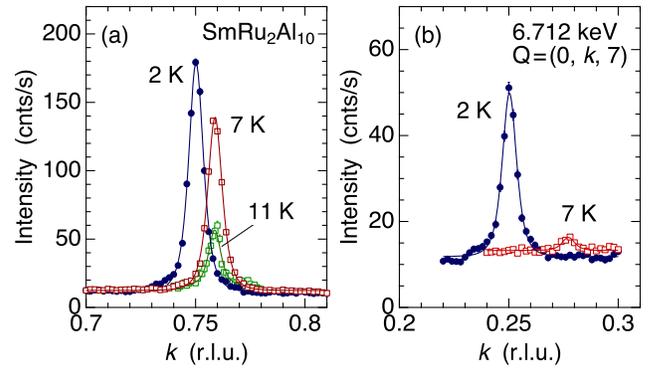}
\end{center}
\caption{(Color online) Scans of the resonant intensity at 6.712 keV in the reciprocal space along $Q=(0, k, 7)$ at 2~K, 7~K, and 11~K, without polarization analysis.  
(a) The main peak at $\bm{q}+\bm{\tau}_{007}=(0, 0.75, 7)$ in the C-phase and $\bm{q}+\bm{\tau}_{007}=(0, 0.759, 7)$ in the IC-phase. 
(b) The third harmonic at $3\bm{q}+\bm{\tau}_{0\bar{2}7}=(0, 0.25, 7)$ in the C-phase and $3\bm{q}+\bm{\tau}_{0\bar{2}7}=(0, 0.278, 7)$ in the IC-phase. }
\label{fig:fig7}
\end{figure}

Figure \ref{fig:fig7} shows the scans of the $E2$ resonant intensity along $(0, k, 7)$ in the reciprocal space. 
The data points were fit with a Lorentzian-squared lineshape. 
As shown in the scan at 7~K in Fig.~\ref{fig:fig7}(a), when the temperature is increased above $T_{\text{M}}$, the peak position shifts slightly from the commensurate position at $k=0.75$ to an incommensurate position at $k=0.759$. 
Therefore, the transition at $T_{\text{M}}$ can be concluded as a commensurate (C) to incommensurate (IC) magnetic phase transition. 
The IC peak position does not change with temperature as shown by the scan at 11~K.

As shown in Fig.~\ref{fig:fig7}(b), a clear peak is observed also in the scan around the third harmonic position at $3\bm{q}+\bm{\tau}_{0\bar{2}7}$ both in the C-phase and the IC-phase. 
A remarkable result is that the intensity of the third harmonic in the IC-phase at 7~K is significantly weaker than that of the main peak at $\bm{q}$, which is in contrast with the intensity ratio in the C-phase at 2~K.

\begin{figure}[t]
\begin{center}
\includegraphics[width=8cm]{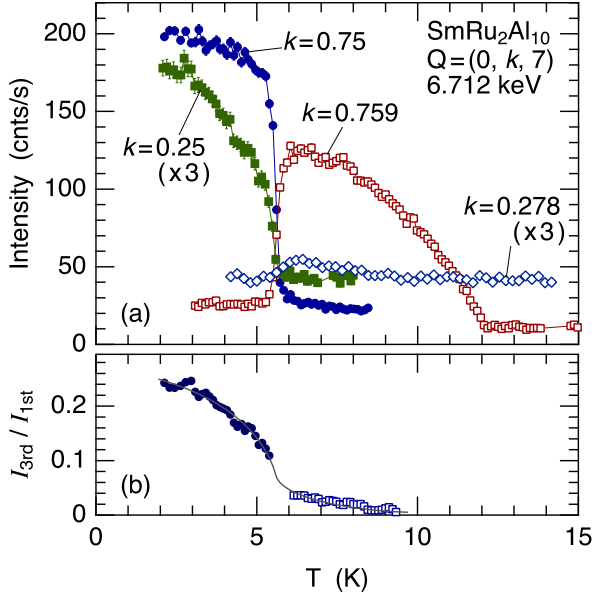}
\end{center}
\caption{(Color online) (a) Temperature dependences of the magnetic Bragg peak intensities at the first and third harmonic positions and at 6.712 keV. 
The intensities of the third harmonic peaks are multiplied by three. 
(b) The ratio of the third harmonic intensity to the first harmonic intensity as a function of temperature after subtraction of the background. Solid line is a guide for the eye. }
\label{fig:fig8}
\end{figure}

Temperature dependences of the first and third harmonic peak intensities are shown in Fig.~\ref{fig:fig8}(a). 
The intensity of the first harmonic at $k=0.75$ in the C-phase interchanges with that of the incommensurate peak at $k=0.759$ at $T_{\text{M}}$, indicating a first order transition. 
The peak at $k=0.759$ decreases continuously with increasing temperature and vanishes at $T_{\text{N}}$. 
By contrast, the intensity of the third harmonic at $k=0.25$ in the C-phase decreases more rapidly than that of the first harmonic. 
Above $T_{\text{M}}$, although the third harmonic peak at $k=0.278$ remains, it is much weaker than that of the first harmonic. 
The intensity ratio of the third harmonic to the first harmonic, after subtraction of the background, is 0.25 at 2~K. 
It decreases with increasing temperature and becomes $\sim0.025$ at 7~K as shown in Fig.~\ref{fig:fig8}(b).

\begin{figure}[t]
\begin{center}
\includegraphics[width=8.5cm]{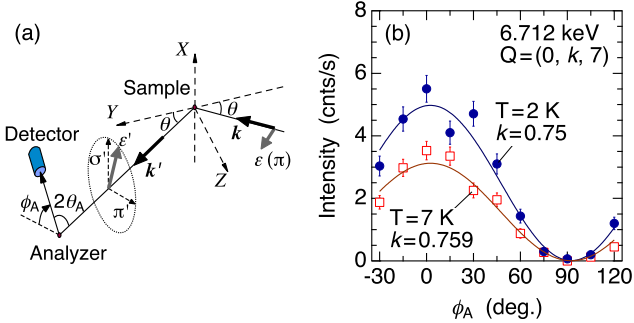}
\end{center}
\caption{(Color online) (a) Scattering geometry of the experiment with polarization analysis. Cu-200 reflection is used as an analyzer, where $2\theta_A=92.55^{\circ}$ at 6.712 keV. 
(b) $\phi_{\text{A}}$ dependence of the resonant intensity at 6.712 keV ($E2$). Solid line is a fit to the data assuming a purely $\pi$-$\sigma'$ scattering. }
\label{fig:fig9}
\end{figure}

Both in the C- and IC-phases, the scattering is purely of $\pi$-$\sigma'$, i.e., the polarization vector of the x-ray rotates from the $\pi$-polarization ($\parallel$ scattering plane) to the $\sigma$-polarization ($\perp$ scattering plane).  
The result of polarization analysis of the $E2$ resonance peak in the C- and IC-phases is shown in Fig.~\ref{fig:fig9}. 
As shown by the solid lines, the data can be well explained by the $\pi$-$\sigma'$ scattering only. 
This result is in good agreement with the interpretation from $\chi_b$ that the antiferromagnetic moment $\bm{\mu}_{\text{AF}}$ is oriented along the $b$ axis both in the C- and IC-phases. 
Also from the result that the $\pi$-$\pi'$ intensity at $\phi_{\text{A}}=90^{\circ}$ vanishes, the possibility of $\bm{\mu}_{\text{AF}}$ being parallel to the $a$ axis ($\parallel\bm{k}\times\bm{k}'$) can be excluded.
These conclusions are deduced from the geometrical factor of Eq.~(\ref{eq:GE2}). 
From the present experiment only, however, we cannot conclude that $\bm{\mu}_{\text{AF}}$ has no $c$ axis component. 
To prove this only from the results of x-ray diffraction, 
it is necessary to measure the azimuthal angle dependence of a magnetic Bragg peak. 
However, from $\chi_b$, $\bm{\mu}_{\text{AF}} \parallel b$ is almost certain and can be used as a reasonable foundation for the following discussion.

\section{Discussion}
\label{sec:Disc}
\subsection{Commensurate phase}
\label{subsec:DiscA}
The present experimental results of RXD can be understood by assuming the same magnetic structure as in NdFe$_2$Al$_{10}$ only by changing the moment direction from $a$ to $b$ axis.~\cite{Robert14} 
It is expressed by the primary wave vector $\bm{q}_{1}=(0, 0.75, 0)$ and its third harmonic $\bm{q}_{3}=(0, 0.25, 0)$. 
This means that the magnetic unit cell is constructed by four chemical unit cells along the $b$ axis. 
From the previous studies of magnetic structure in RT$_2$Al$_{10}$, it is almost certain that the nearest neighbor moments along the $c$ axis are strongly coupled by an antiferromagnetic exchange interaction $J_1$. 
Then, there arise two types of antiferromagnetic arrangements with respect to this $J_1$ pair. 
One is the $(+-)$-type, named A, where the moment at $(0, y, 1/4)$ and $(0, -y, 3/4)$ is oriented to $+$ and $-$ direction, respectively. 
The other is the $(-+)$-type, named B, where the moments are oppositely oriented. 
Among the $2^8=256$ kinds of arrangements in the magnetic unit cell, only two types of arrangements, AAAABBBB and ABABBABA, with any cyclic transformation allowed, give rise to $\bm{q}_{1}$ and $\bm{q}_{3}$ components only. 
The intensity ratio $I_{\text{3rd}}/I_{\text{1st}}$ in Fig.~\ref{fig:fig8} is consistent with the magnetic structure factor of the ABABBABA arrangement, which is the same as that of NdFe$_2$Al$_{10}$.~\cite{Robert14} 

\begin{figure}[t]
\begin{center}
\includegraphics[width=8.5cm]{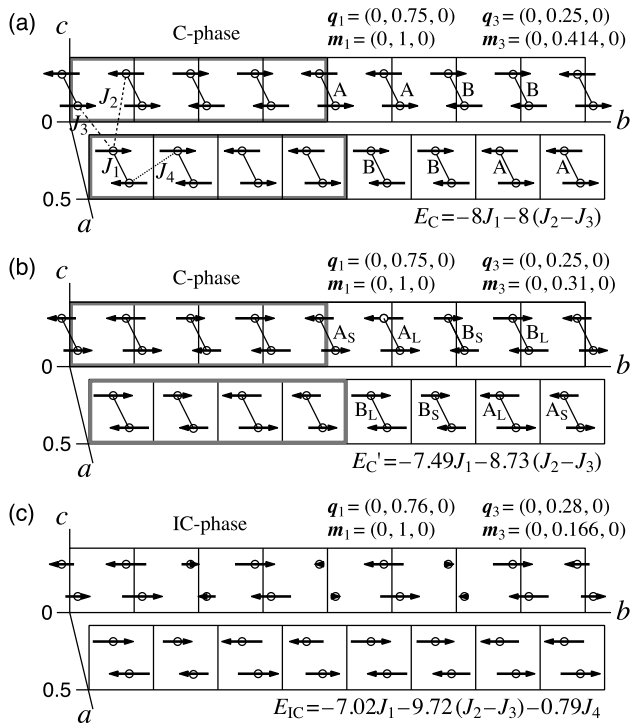}
\end{center}
\caption{Schematic representation of the magnetic structure models corresponding to the (a) C-phase at 2~K, (b) C-phase at 5~K, and (c) IC-phase at 7~K. 
The Sm atoms on the $bc$-plane with $x=0$ and $x=0.5$ are shown. The magnetic unit cell in the C-phase is indicated by thick rectangles, in which the ABABBABA arrangement is displayed. 
The subscripts S and L represents the short and long moment pair, respectively.
Magnetic moment pairs up to fourth neighbor sites are represented by solid, dashed, dot-dashed, and dotted lines, which corresponds to $J_1$, $J_2$, $J_3$, and $J_4$, respectively. 
}
\label{fig:fig10}
\end{figure}

In the ABABBABA arrangement, the Sm moments at the $j$th lattice point, $\bm{\mu}_{1/4,j}$ and $\bm{\mu}_{3/4,j}$ on the $z=1/4$ and $z=3/4$ layer, respectively, are written as
\begin{align}
\bm{\mu}_{1/4,j} &= \bm{m}_{1} \cos 2\pi (\bm{q}_1 \cdot \bm{r}_{j} + 3/16)  \nonumber \\
                         &\;\;\;\; +\bm{m}_{3} \cos 2\pi (\bm{q}_3 \cdot \bm{r}_{j} + 1/16) \label{eq:M1} \\
\bm{\mu}_{3/4,j} &= \bm{m}_{1} \cos 2\pi (\bm{q}_1 \cdot \bm{r}_{j} - 5/16)  \nonumber \\
                         &\;\;\;\; +\bm{m}_{3} \cos 2\pi (\bm{q}_3 \cdot \bm{r}_{j} - 7/16) \label{eq:M2}                
\end{align}
where $\bm{m}_{1}$ and $\bm{m}_{3}$ are the magnetic amplitude vectors for $\bm{q}_{1}$ and $\bm{q}_{3}$, respectively. 
$\bm{r}_j$ is the position of the $j$th lattice point. 
By setting $\bm{m}_{1}=(0,1,0)$ and $\bm{m}_{3}=(0,\sqrt{2}-1,0)$, and by selecting the phase factors as above, the ABABBABA arrangement of the $J_1$ pairs is obtained, where all the magnitudes of Sm moments are equal.\cite{note1} 
This magnetic structure is shown in Fig.~\ref{fig:fig10}(a). 
Using this model, we can calculate the intensity by the relation $I(\bm{Q})\propto |F(\bm{Q})|^2$, where 
\begin{equation}
F(\bm{Q})=\sum_{j} f_j \exp (i \bm{Q} \cdot \bm{r}_j)
\end{equation}
is the structure factor of the Bragg peak. The atomic scattering factor of the $j$-th Sm ion, $f_j$, is calculated by using Eq.~(\ref{eq:E2}) for the $E2$ resonance. 
Although the spectral function $\alpha_{E2}(\omega)$ is unknown, it cancels out in the intensity ratio of $I_{\text{3rd}}/I_{\text{1st}}$. 
The calculated ratio, $I_{\text{3rd}}/I_{\text{1st}}=0.21$, agrees well with the experimental value at 2~K in Fig.~\ref{fig:fig8}(b).

The calculated value of $I_{\text{3rd}}/I_{\text{1st}}$ for the $E1$ resonance is 0.28. 
This value also agrees with the data shown in Fig.~\ref{fig:fig6} within the experimental accuracy, which is worse than that of the $E2$ resonance due to the weak intensity of $I_{\text{3rd}}$ as can be seen in Fig.~\ref{fig:fig5}(b). 
As for the nonresonant scattering, it is necessary to separate the magnetic moment into orbital and spin parts to calculate the scattering amplitude. 
Since we have no such information in SmRu$_{2}$Al$_{10}$, we simply assume $\mu_{l}=(1-g/2)J$ and $\mu_{s}=(g-1)J$, where $g=2/7$ and $J=5/2$ for Sm$^{3+}$. 
Then, using Eq.~(\ref{eq:nrm}), $I_{\text{3rd}}/I_{\text{1st}}$ is calculated to be 0.26, which agrees with the data in Fig.~\ref{fig:fig6}. 
In actual fact, all these values of $I_{\text{3rd}}/I_{\text{1st}}$ are almost equal to $|F(\bm{Q}_3)|^2/|F(\bm{Q}_1)|^2=0.24$, where $F(\bm{Q})=\sum_{j} \mu_{b,j} \exp (i\bm{Q}\cdot\bm{r}_j)$ is the magnetic structure factor, $\bm{Q}_1=(0, 0.75, 7)$, and $\bm{Q}_3=(0, 0.25, 7)$. 
The difference of geometrical factors between $\bm{Q}_1$ and $\bm{Q}_3$ gives only a slight modification. 
To conclude, the magnetic structure model of Fig.~\ref{fig:fig10}(a) is consistent with the result of Fig.~\ref{fig:fig6} that $I_{\text{3rd}}/I_{\text{1st}}$ is approximately 0.25 over the whole energy range from 6.68 keV to 6.72 keV. 
To be more reliable, of course, it is necessary to collect the intensities of other magnetic Bragg peaks, including also the energy dependences, and examine the consistency of the model structure.

The intensity ratio $I_{\text{3rd}}/I_{\text{1st}}$ decreases with increasing temperature as shown in Fig.~\ref{fig:fig8}(b). 
This means that $m_3$ decreases more rapidly than $m_1$. 
We show in Fig.~\ref{fig:fig10}(b) a model of the magnetic structure at elevated temperatures in the C-phase, 
where $m_3$ is set to be 0.31, which is 0.75 times the full amplitude of $\sqrt{2}-1$. 
The calculated $I_{\text{3rd}}/I_{\text{1st}}$ is 0.12, which corresponds to the temperature around 5~K from Fig.~\ref{fig:fig8}(b). 
In this structure, there appear $J_1$ pairs with long and short magnetic moments. 
Their lengths are 1.05 and 0.87 times the moment value in Fig.~\ref{fig:fig10}(a). 

There is also a possibility of changing the phase parameters in Eqs.~(\ref{eq:M1}) and (\ref{eq:M2}) because the intensity is not affected by the phase. 
However, if we change the phase parameters, the averaged magnitude of the ordered moments become smaller than that of Fig.~\ref{fig:fig10}(b). 
The phase parameters in Eqs.~(\ref{eq:M1}) and (\ref{eq:M2}) give the maximum value of the averaged magnetic moment for a given amplitude of $m_3$. 
We consider that this is a reasonable assumption from the viewpoint of exchange energy and magnetic entropy.

\subsection{Incommensurate phase}
\label{subsec:DiscB}
By changing the propagation vector to $\bm{q}_{1}=(0, 0.76, 0)$ and $\bm{q}_{3}=(0, 0.28, 0)$, 
and by using a decreased amplitude of $\bm{m}_{3}=(0, 0.166, 0)$, 
an incommensurate magnetic structure as shown in Fig.~\ref{fig:fig10}(c) is obtained, where the the first eight unit cells are depicted.\cite{note2} 
The intensity ratio $I_{\text{3rd}}/I_{\text{1st}}$ for the $E2$ resonance is calculated to be 0.03, which agrees well with the experimental value at 7~K in Fig.~\ref{fig:fig8}(b). 
This is more like a sinusoidal structure than the squared one in the C-phase. 
The magnitude of the moments oscillates. However, there still remain a squared feature of the ABABBABA arrangement, 
which corresponds to the fact that the third harmonic intensity still remains in the IC-phase. 

In Fig.~\ref{fig:fig10}(c), we see some Sm sites where the moment values are very small. 
These small moment sites alternately appear also in the $x=0.5$ layer, although they are not displayed in Fig.~\ref{fig:fig10}(c) because they are outside the range of the figure. 
Although Fig.~\ref{fig:fig10}(c) presumes a static order, there is no evidence to consider the moments are statically fixed.  
There is also a possibility that Fig.~\ref{fig:fig10}(c) can be a snap shot of a propagating IC-wave. 

Another possibility for the origin of the Bragg peak at $(0, 0.76, 0)$ is a regular occurrence of discommensuration. 
The commensurate ABABBABA arrangement contains two discommensurations, i.e., AA and BB arrangements, in four unit cells. 
This gives Bragg peaks at $(0, 0.75, 0)$ and at $(0, 0.25, 0)$. 
If there occurs another discommensuration in every 25 unit cells and returns to the original phase in every 100 unit cells, 
for example, we have a first harmonic peak at $(0, 0.76, 0)$. 
However, in such a case, the third harmonic peak will also be shifted by 0.01, not by 0.03 as observed in the experiment. 
Furthermore, side peaks are expected to appear at $(0, 0.76\pm 0.04, 0)$, 
which are not detected above the background in the present experiment as shown in Fig.~\ref{fig:fig7}.  
Therefore, the possibility of regular occurrence of another discommensuration can be discarded. 

\subsection{Magnetic exchange energy}
\label{subsec:DiscC}
As performed in Ref.~\onlinecite{Robert14}, let us calculate and compare the exchange energies of the C- and IC-phase. 
We take 100 unit cells, corresponding to the magnetic unit cell of the IC-structure with $\bm{q}_{1}=(0, 0.76, 0)$, calculate the total exchange energy, and convert it to the energy per four unit cells, corresponding to the magnetic unit cell of the C-phase. 
In the C-phase, the energy estimated from the structure of Fig.~\ref{fig:fig10}(a) is $E_{\text{C}}= - 8J_1 - 8(J_2 - J_3)$ per magnetic unit cell, where $J_i$ is taken positive for an antiferromagnetic pair. The $J_4$ term does not contribute to the total energy. 
If we assume the ABABABAB-structure, which is realized in CeRu$_2$Al$_{10}$ without discommensuration and is described by $\bm{q}=(0,1,0)$, 
the energy becomes $E_{0}=-8J_1 -16 (J_2 - J_3) -16J_4$. 
Without the $J_4$ term, $J_2 < J_3$ is necessary so that $E_{\text{C}} < E_{0}$ is fulfilled, as discussed in Ref.~\onlinecite{Robert14}. 
What is contradictory is that, if $J_2 < J_3$ were the case, $E_{\text{C}}$ cannot be the minimum energy among other possible structures such as AAAABBBB, where the energy is $- 8J_1 + 8 (J_2 - J_3)$. 
However, by including the $J_4$ term, $E_{\text{C}} < E_{0}$ can be realized even if $J_2 > J_3$, if $2J_4 < - (J_2 - J_3)$ is fulfilled (ferromagnetic $J_4$). 
This shows that the long-range interaction of $J_4$ should be taken into account to explain the discommensuration in the ABABBABA arrangement. 

In the magnetic structure of Fig.~\ref{fig:fig10}(b), representing the C-structure at an elevated temperature of 5~K, 
the energy is calculated to be $E_{\text{C}}'= - 7.49J_1 - 8.73(J_2 - J_3)$. 
This shows that the energy gain from the $J_1$ term decreases, whereas that from the $(J_2 - J_3)$ term increases. 

In the IC-phase, the energy becomes $E_{\text{IC}}=-7.02 J_1 - 9.72 (J_2 - J_3) - 0.79 J_4$. 
This again shows that, by the phase transition from C to IC, the $J_1$ term gains less energy while the $(J_2 - J_3)$ term gains more energy. The $J_4$ term has only a marginal effect. 
Therefore, the appearance of the IC-structure at high temperatures, as well as the decrease of the third harmonic amplitude in the C-phase with increasing temperature, may be regarded as reflecting the RKKY magnetic exchange interactions, where more energy gain is obtained from the long-range interactions of $J_2$ and $J_3$ terms. 
At low temperatures, to reduce magnetic entropy at the small moment sites, and also to gain more magneto-elastic coupling energy by increasing the magnitude of the magnetic moment, the commensurate structure with equal moment values is expected to be realized. 

\subsection{Crystal field analysis of magnetic susceptibility}
\label{subsec:DiscD}
To discuss the anisotropy of magnetic susceptibility in Fig.~\ref{fig:fig3}, we consider the following mean-field Hamiltonian: 
\begin{equation}
\mathcal{H}=\mathcal{H}_{\text{SO}} + \mathcal{H}_{\text{CEF}}  -\bm{\mu}\cdot\bm{H} 
-J_{\text{ex}}\bm{\mu}\cdot \langle \bm{\mu} \rangle\,,
\label{eq:CEF}
\end{equation}
where the first, second, third, and fourth term represent spin-orbit interaction, CEF, Zeeman energy, and magnetic exchange energy in a two-sublattice model, respectively. 
The ground multiplet of $\mathcal{H}_{\text{SO}}$ for Sm$^{3+}$ ($4f^5$: $L=5$ and $S=5/2$) is represented by $J=5/2$. 
In addition, we also consider the $J=7/2$ first excited state, which we assume to be located at 1500~K.\cite{Alekseev97} 
Diagonalization of the Hamiltonian is performed in the 14-dimensional space consisting of $J=5/2$ and $J=7/2$ multiplet states. 
A method of calculating the matrix elements is described in the Appendix \ref{app:B}.  

By taking into account the $\hat{C}^{(2)}_{\pm 2}$ term only $(q_{2,\pm 2}\langle r^2 \rangle = 5500$~K), corresponding to $B_{22}=9.0$~K in the conventional expression, 
we have three Kramers doublets split from the $J=5/2$ ground multiplet; the level scheme is 0~K -- 47~K -- 95~K.  
The ground state has large magnetic moments of $\pm 0.69$~$\mu_{\text{B}}$ ($\sim gJ$) along the $b$ axis, whereas those along the $a$ and $c$ axes are less than 0.062~$\mu_{\text{B}}$. 
In reverse, the magnetic moments of the excited state at 95~K are $\pm 0.69$~$\mu_{\text{B}}$ along the $a$ axis. 
These are consistent with the experimental result of $\chi_b > \chi_c > \chi_a$. 
If we put $J_{\text{ex}}=27$~K/$\mu_{\text{B}}^{\;2}$ as an exchange parameter, we obtain an antiferromagnetic ordered state with an ordered moment of $\mu_{\text{AF}}$=0.68~$\mu_{\text{B}}$ oriented along the $b$ axis and a transition temperature of $T_{\text{N}}=12$~K. 
On the other hand, if we assume an ordered state with the moments oriented along the $a$ or $c$ axis, the resultant values of $\mu_{\text{AF}}$ and $T_{\text{N}}$ become extremely small, which are less than 0.1~$\mu_{\text{B}}$ and 0.2~K, respectively. 
This result reflects the magnetic anisotropy of the CEF ground state. 

\begin{figure}[t]
\begin{center}
\includegraphics[width=8.5cm]{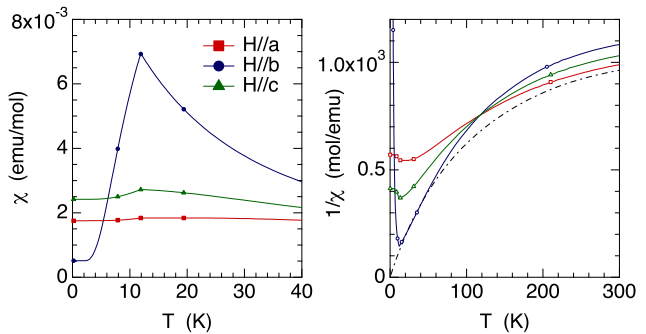}
\end{center}
\caption{(Color online) Calculated temperature dependences of magnetic susceptibility and inverse magnetic susceptibility using a mean-field model and crystal field parameters described in the text. The dot-dashed curve shows $1/\chi$ for a free Sm$^{3+}$ ion with $J=5/2$ ground state and $J=7/2$ excited state at 1500~K. 
}
\label{fig:fig11}
\end{figure}

\begin{figure}[t]
\begin{center}
\includegraphics[width=8.5cm]{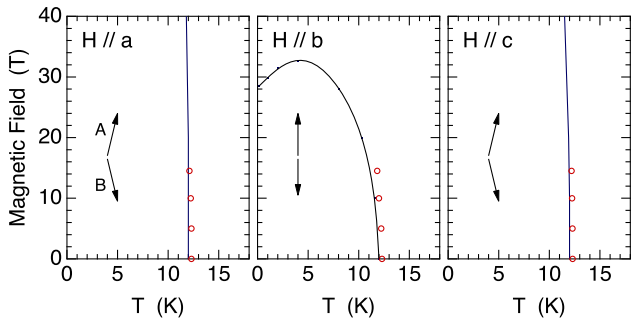}
\end{center}
\caption{(Color online) Calculated magnetic phase diagram for the two-sublattice mean-field Hamiltonian described in the text. 
The open circles represent the experimental phase boundary determined from $\rho(T,H)$ in Fig.~\ref{fig:fig2}. 
Arrows represent the schematic of the ordered moments in the A- and B-sublattices. 
}
\label{fig:fig12}
\end{figure}

The calculated magnetic susceptibilities for the three field directions obtained from the above mean-field Hamiltonian are shown in Fig.~\ref{fig:fig11}. 
The magnetic anisotropy at low temperatures is well reproduced. 
Since we considered only the $\hat{C}^{(2)}_{\pm 2}$ term, and have not tried to fit the data by including more terms in $\mathcal{H}_{\text{CEF}}$, there still remains a room for further improvement in the fitting, such as the absolute values of $\chi(T)$ at low temperatures. 
However, in the present two-sublattice Hamiltonian, it seems difficult to reproduce the increase of $\chi_a(T)$ and $\chi_c(T)$ with decreasing $T$ even below $T_{\text{N}}$ as shown in Fig.~\ref{fig:fig3}. 
The calculated susceptibilities of $\chi_a(T)$ and $\chi_c(T)$ both exhibit a cusp at $T_{\text{N}}$ and decreases with decreasing $T$.  

Finally, the calculated magnetic phase diagram for the present mean-field Hamiltonian is shown in Fig.~\ref{fig:fig12}. 
The almost field independent $T_{\text{N}}$ for $H\parallel a$ and $H\parallel c$, and the slight decrease of $T_{\text{N}}$ at high fields for $H\parallel b$, is well reproduced. The critical field for $H \perp \mu_{\text{AF}}$ is extremely high; it is 470 T for $H\parallel a$ and 147 T for for $H\parallel c$ at 0~K. 
This is due to the crystal field anisotropy, which prevents the magnetic moment to be oriented to the $a$ and $c$ axes. 
The moments are only weakly canted to the field direction, as schematically illustrated in the inset figures. 
On the other hand, the critical field of 28 T at 0~K for $H \parallel b \parallel \mu_{\text{AF}}$ is determined by the condition that the Zeeman energy $\mu H$ becomes equal to the exchange energy $J_{\text{ex}}\mu^{2}$, where $\mu=0.68$ $\mu_{\text{B}}$. 
The critical field increases to $\sim 33$ T at 4~K because the magnetic moment in the B-sublattice, which is directed opposite the magnetic field, is shortened at finite temperatures and can suppress the loss of Zeeman energy. 
The calculation shows that the field-insensitive $T_{\text{N}}$ in this compound reflects the small $g$-factor of $2/7$ for Sm$^{3+}$ and the strong crystal filed anisotropy.

\section{Conclusions}
We have performed electrical resistivity, magnetic susceptibility, specific heat, and resonant x-ray diffraction measurements on SmRu$_2$Al$_{10}$ using a single crystal. 
Successive phase transitions have been confirmed to take place at $T_{\text{N}}$ = 12.3~K and at $T_{\text{M}}$ = 5.6~K. 
The magnetization easy axis in the paramagnetic phase is the $b$ axis and the ordered moment is also oriented along the $b$ axis. 
The single ion crystal-field anisotropy and the direction of the ordered moment are consistent with each other and can be understood in the framework of the normal crystal field theory. This point is different from the case in CeRu$_{2}$Al$_{10}$, where the direction of the ordered moment is different from the crystal-field easy axis.

The magnetic structure at the lowest temperature is described by $\bm{q}_1=(0, 0.75, 0)$ and $\bm{q}_3=(0, 0.25, 0)=3\bm{q}_1 + \bm{\tau}_{0\bar{2}0}$, where four chemical unit cells form the magnetic unit cell. 
We have proposed a magnetic structure, which is identical to that of NdFe$_{2}$Al$_{10}$ except the direction of the moment.
Above $T_{\text{M}}$, the commensurate structure changes to an incommensurate one, which is described by $\bm{q}_1=(0, 0.759, 0)$ and $\bm{q}_3=(0, 0.278, 0)=3\bm{q}_1 + \bm{\tau}_{0\bar{2}0}$. The intensity of the third harmonic peak exhibits a significant decrease in the IC phase. 
From a simple energy estimate, the C-IC transition is suggested to reflect the long-range magnetic exchange interaction of RKKY type.

\section*{Acknowledgements}
The authors thank Dr. J.-M. Mignot for valuable discussions. 
This work was supported by Grants-in-Aid for Scientific Research (Nos. 24540376, 24340087, 15K05173, and 15K05175) from the Japan Society for the Promotion of Science (JSPS). 
The synchrotron experiments were performed under the approval of the Photon Factory Program Advisory Committee (No. 2014G-035). 

\appendix
\section{resonant and nonresonant atomic scattering factors from magnetic dipole moment}
\label{app:A}
The process of resonant scattering by an atom with a magnetic moment $\bm{\mu}$ is called \textit{resonant exchange scattering}.\cite{Hannon88} 
This term means that the resonant signal reflects the electronic state of the unoccupied shell involved in the resonance, which is modified by the exchange interaction with the local electrons responsible for the magnetic moment. 
Therefore, in a strict sense, resonant scattering does not directly observe the order parameter itself. 
However, it is generally accepted that the scattering amplitude is proportional to and contains much information of the order parameter.  

The atomic scattering factors for $E1$ and $E2$ resonances originating from a magnetic dipole moment $\bm{\mu}$ in the localized orbital is expressed as 
\begin{align}
f_{E1}(\omega) &= \alpha_{E1}(\omega)\bm{G}_{E1} \cdot \bm{\mu} \label{eq:E1} \,, \\
f_{E2}(\omega) &= \alpha_{E2}(\omega)\bm{G}_{E2} \cdot \bm{\mu} \label{eq:E2} \,,
\end{align}
where $\bm{G}$ and $\alpha(\omega)$ represent the geometrical factor and the spectral function, respectively, for the magnetic dipole order parameter (rank-1 tensor).\cite{Nagao06,Nagao08,Nagao10} 

The rank-1 geometrical factors are expressed as 
\begin{align}
\bm{G}_{E1}  &= \bm{\varepsilon}' \times \bm{\varepsilon} \label{eq:GE1} \,, \\ 
\bm{G}_{E2}  &= \bigl\{(\bm{k}' \cdot \bm{k})(\bm{\varepsilon}' \times \bm{\varepsilon})
+(\bm{\varepsilon}' \cdot \bm{\varepsilon})(\bm{k}' \times \bm{k}) \nonumber \\
 &\;\;\;\; +(\bm{k}' \cdot \bm{\varepsilon})(\bm{\varepsilon}' \times \bm{k})
+(\bm{\varepsilon}' \cdot \bm{k})(\bm{k}' \times \bm{\varepsilon}) \bigr\} \label{eq:GE2} \,,
\end{align}
where $\bm{k}$ ($\bm{k'}$) and $\bm{\varepsilon}$ ($\bm{\varepsilon'}$) represent the wave vector and polarization vector of the incident (final) x-ray, respectively. 
The $xyz$-coordinates are taken so that 
$\hat{\bm{x}}\parallel \bm{k}\times\bm{k}'$, $\hat{\bm{y}}\parallel\bm{k}+\bm{k}'$, and $\hat{\bm{z}} \parallel \bm{Q}=\bm{k}-\bm{k}'$. 
In the present scattering geometry of Fig.~\ref{fig:fig9}(a), $\bm{\varepsilon}$ is in the $yz$ scattering plane ($\pi$-polarized), which coincides with the $bc$ plane of the crystal. 
Since $\bm{\mu}$ is expected to be parallel to the $b$ axis, only the scattering factor for $\pi$-$\sigma'$ ($\bm{\varepsilon}' \parallel \hat{\bm{x}}$) remains nonzero in both $E1$ and $E2$ resonances. 

With respect to the spectral functions in Eqs.~(\ref{eq:E1}) and (\ref{eq:E2}), which are actually unknown factors, 
it is convenient to use the following form:
\begin{equation}
\alpha(\omega)=\frac{i C \Gamma }{\hbar\omega - \Delta + i\Gamma} \exp(i\phi)\,,
\label{eq:alpha}
\end{equation}
where $\Delta$, $\Gamma$, and $C$ represent the energy, width, and magnitude of the resonance peak, respectively.  
We introduce $\phi$ as an empirical phase parameter to express the real and imaginary parts of $\alpha(\omega)$. 
Note that another subscript, $E1$ or $E2$, is attached to these parameters.

Nonresonant magnetic scattering of x-rays by a magnetic dipole moment is caused by a direct interaction between x-ray photon and the localized electron.\cite{Lovesey96} The atomic scattering factor is expressed as 
\begin{equation}
f_{\text{nrm}}(\omega) = i (\frac{e^2}{mc^2}) (\frac{\hbar\omega}{mc^2})
\left( \bm{\mu}_{l} \cdot \bm{A}+\bm{\mu}_{s} \cdot \bm{B} \right)\,,
    \label{eq:nrm}
\end{equation}
where $\bm{\mu}_{l}$ and $\bm{\mu}_{s}$ are the orbital and spin part of the magnetic form factor, respectively. 
$e^2/mc^2=2.82 \times 10^{-13}$ cm is the classical radius of an electron, and $mc^2$ corresponds to 511 keV. 
The geometrical factors of $\bm{A}$ and $\bm{B}$ are expressed as 
\begin{align}
\bm{A} &= -2(1-\hat{\bm{k}}\cdot\hat{\bm{k}}')\bigl[\hat{\bm{Q}}\times\{
   (\bm{\varepsilon}'\times\bm{\varepsilon})\times\hat{\bm{Q}}\} \bigr] \,, \label{eq:facA} \\
    \bm{B} &= \bigl\{ (\bm{\varepsilon}'\times\bm{\varepsilon}) 
           -(\bm{\varepsilon}\cdot\hat{\bm{k}}')(\bm{\varepsilon}'\times\hat{\bm{k}}')  \nonumber \\
    & +(\bm{\varepsilon}'\cdot\hat{\bm{k}})(\bm{\varepsilon}\times\hat{\bm{k}})
           -(\hat{\bm{k}}'\times\bm{\varepsilon}')\times(\hat{\bm{k}}\times\bm{\varepsilon}) \bigr\}\;. \label{eq:facB}
\end{align}
The orbital and spin parts of the magnetic form factor are defined as
\begin{align}
  \bm{\mu}_l &= -\frac{1}{2\mu_{\text{B}}}\int \bm{\mu}_l(\bm{r}) e^{i\bm{Q}\cdot\bm{r}} d\bm{r}\,, \label{eq:mul} \\
  \bm{\mu}_s &= -\frac{1}{2\mu_{\text{B}}}\int \bm{\mu}_s(\bm{r}) e^{i\bm{Q}\cdot\bm{r}} d\bm{r} \label{eq:mus} \,,
\end{align}
which depend on the scattering vector $\bm{Q}=\bm{k} - \bm{k}'$. 

An advantage of nonresonant magnetic scattering is that the absolute value of the magnetic moment is directly related with the scattering amplitude. However, in normal experimental cases, we have little reliable information about the separation of $\bm{\mu}$ into $\bm{\mu}_{l}$ and $\bm{\mu}_{s}$. In the discussion in this paper, we simply assumed 
$\bm{\mu}_{l}=(1-g/2)\bm{J}$ and $\bm{\mu}_{s}=(g-1)\bm{J}$. 
Another difficulty is that the observed intensity is generally much ($5\sim 7$ orders of magnitude) weaker than that of the charge (Thomson) scattering from the fundamental lattice and it is difficult to find the reliable scale factor to estimate the absolute value of the magnetic moment. 

The observed intensity $I(\bm{Q},\omega)$ is proportional to the square of the absolute value of the total structure factor, which is expressed as 
\begin{equation}
I(\bm{Q},\omega) \propto \Bigl| \sum_{j} \{ f_{\text{nrm}}(\omega) + f_{E2}(\omega)+ f_{E1}(\omega) \}_j \exp (i\bm{Q}\cdot\bm{r}_j) \Bigr|^2 ,
\label{eq:Int}
\end{equation}
Using Eq.~(\ref{eq:Int}) and the ABABBABA-type magnetic structure model described in the text, 
we can fit the energy dependences of the intensity shown in Fig.~\ref{fig:fig6} by treating the parameters in $\alpha_{E2}(\omega)$ and $\alpha_{E1}(\omega)$ as free parameters. 
The fitting shows that the three terms in Eq.~(\ref{eq:Int}) interfere with each other. 
However, we could not uniquely determine the phase parameters of $\phi_{E2}$ and $\phi_{E1}$, and this analysis did not provide more useful information than those described in Sec.~\ref{subsec:DiscA}. 
Detailed investigation of several other magnetic Bragg peaks is necessary to extract reliable phase parameters of the spectral functions.

\section{Crystal-Field Hamiltonian }
\label{app:B}
To calculate the magnetic susceptibility of a rare-earth ion in a CEF potential by taking into account the excited $J$-multiplets, 
it is necessary to deduce the reduced matrix element, or the so-called Stevens' factor, for the multipole operators involved in the CEF Hamiltonian. 
We cannot neglect contributions from the excited $J$-multiplet in describing the physical properties of Sm$^{3+}$ ions, in which 
the $J=7/2$ excited level is located at around 1500~K.  

First, we expand the CEF potential by using spherical harmonics.\cite{Stevens52,Hutchings64} 
In a point-charge-model, the CEF for an electron at position $(r, \theta, \phi)$ is expressed as 
\begin{align}
V_{\text{c}}(r, \theta, \phi) &= \sum_{l=0}^{\infty} \sum_{m=-l}^{l}  \sum_{j=1}^{N} r^l \left( \frac{4\pi}{2l + 1} \right) \left(\frac{-Z_j e^2}{R_j ^{\; l+1}} \right) \nonumber \\
&\;\;\;\; \times Y_{m}^{(l) *} (\theta_j, \phi_j) Y_{m}^{(l)}(\theta, \phi) \nonumber \\
&= \sum_{l=0}^{\infty} \sum_{m=-l}^{l} r^l q_{l,m} C^{(l)}_m (\theta, \phi) \, ,
\label{eq:CEFV}
\end{align}
where $Z_j$ and $(R_j, \theta_j, \phi_j)$ represent the effective charge and its position parameters, respectively, of the $j$th surrounding ion. 
$C^{(l)}_m$ is defined as $\sqrt{4\pi/(2l + 1)} Y^{(l)}_m$. 
The CEF parameter in this formalism is 
\begin{equation}
q_{l,m} = \sqrt{\frac{4\pi}{2l + 1}} \sum_{j=1}^{N} \left(\frac{-Z_j e^2}{R_j ^{\; l+1}} \right) Y_{m}^{(l)*} (\theta_j, \phi_j)\;.
\label{eq:qlm}
\end{equation}

Using the Wigner-Eckart theorem, the matrix element of the CEF Hamiltonian can be expressed as 
\begin{align}
&\langle J, M | \mathcal{H}_{\text{CEF}} | J', M' \rangle \nonumber \\
&= \sum_{l=0}^{6} \sum_{m=-l}^{l} q_{l,m} \langle r^l \rangle \langle J M | \sum_{i=1}^n C^{(l)}_m(\theta_i,\phi_i) | J' M' \rangle \\
&= \sum_{l=0}^{6} \sum_{m=-l}^{l} q_{l,m} \langle r^{l} \rangle (J||\hat{C}^{(l)}||J') \frac{\langle JM | J'M'lm \rangle}{\sqrt{2J+1}}  \;,
\label{eq:CEFelem}
\end{align}
where $(J||\hat{C}^{(l)}||J')$ represents the reduced matrix element of $\hat{C}^{(l)}_m\equiv \sum_{i=1}^n C^{(l)}_m(\theta_i,\phi_i)$, $\langle JM | J'M'lm \rangle$ the Clebsch-Gordan coefficient, and the sum over $l$ is taken for even integers. 
The reduced matrix elements for the ground and first excited $J$-multiplets of Sm$^{3+}$ is summarized in Table~\ref{tbl:1}, 
which was calculated by following the procedure explained in Ref.~\onlinecite{Stevens52}.  
\begin{table}
\caption{Reduced matrix elements of $(J || C^{(l)} || J')$ for Sm$^{3+}$ ($4f^{5}$, $L=5$, $S=5/2$). $J_0=5/2$ and $J_1=7/2$. }
\label{tbl:1}
\begin{ruledtabular}
\begin{tabular}{ccccc}
$l$ & $(J_0 || C^{(l)} || J_0)$ & $(J_1 || C^{(l)} || J_1)$ & $(J_0 || C^{(l)} || J_1)$ & $(J_1 || C^{(l)} || J_0)$ \\ 
\colrule
2 & $\displaystyle{ \frac{26}{945}}$ & $\displaystyle{\frac{26}{1575} \sqrt{2}}$ & $\displaystyle{\frac{26}{567 \sqrt{5}}}$ & $\displaystyle{-\frac{26}{567\sqrt{5}}}$ \\
4 & $\displaystyle{ \frac{26}{10395} \sqrt{\frac{2}{11}}}$ & $\displaystyle{-\frac{416}{1029105}}$ & $\displaystyle{\frac{52}{22869} \sqrt{\frac{2}{3}}}$ & $\displaystyle{-\frac{52}{22869} \sqrt{\frac{2}{3}}}$ \\
6 & $\displaystyle{ 0}$ & $\displaystyle{\frac{136}{127413} \sqrt{\frac{2}{7}}}$ & $\displaystyle{-\frac{136}{382239} \sqrt{\frac{5}{7}}}$ & $\displaystyle{\frac{136}{382239} \sqrt{\frac{5}{7}}}$ \\
\end{tabular}
\end{ruledtabular}
\end{table}

\begin{table}
\caption{Reduced matrix elements of $(J || \hat{\mu}^{(1)} || J')$ for Sm$^{3+}$ ($4f^{5}$, $L=5$, $S=5/2$). $J_0=5/2$ and $J_1=7/2$. }
\label{tbl:2}
\begin{ruledtabular}
\begin{tabular}{ccccc}
 & $(J_0 || \hat{\mu} || J_0)$ & $(J_1 || \hat{\mu} || J_1)$ & $(J_0 || \hat{\mu} || J_1)$ & $(J_1 || \hat{\mu} || J_0)$ \\ 
\colrule
 & $\displaystyle{ \sqrt{\frac{30}{7}}}$ & $\displaystyle{\frac{52}{3} \sqrt{\frac{2}{7}}}$ & $\displaystyle{-6 \sqrt{\frac{5}{7}}}$ & $\displaystyle{6\sqrt{\frac{5}{7}}}$ \\
\end{tabular}
\end{ruledtabular}
\end{table}

The matrix elements of the magnetic dipole moment, $\bm{\mu}=\bm{L}+2\bm{S}$, is calculated in the same manner by using the characteristic that it is the rank-1 tensor:
\begin{equation}
\langle J, M | \hat{\mu}^{(1)}_{m} | J', M' \rangle = (J||\hat{\mu}^{(1)}||J') \frac{\langle JM | JM'1m \rangle}{\sqrt{2J+1}} \,,
\end{equation}
where $(J||\hat{\mu}^{(1)}||J')$ represents the reduced matrix element of the rank-1 spherical tensor $\hat{\mu}^{(1)}_m$. 
The reduced matrix elements are summarized in Table.~\ref{tbl:2}. 
The spherical operator $\hat{\mu} ^{(1)}_{m}$ is related with $(\mu_x, \mu_y, \mu_z)$ as follows: 
\begin{align}
\mu_x &= \frac{1}{\sqrt{2}} (-\hat{\mu}^{(1)}_{1} + \hat{\mu}^{(1)}_{-1})\,, \nonumber \\
\mu_y &= \frac{i}{\sqrt{2}} (\hat{\mu}^{(1)}_{1} + \hat{\mu}^{(1)}_{-1})\,,  \\
\mu_z &= \hat{\mu}^{(1)}_{0} \nonumber\,.
\end{align}


\begin{thebibliography}{99}
\bibitem{note1} The difference of the phase factors from those of Ref.~\onlinecite{Robert14} is due to the different choice of the origin.
\bibitem{note2} We use $0.76$ instead of 0.759 to simplify the discussion.
\end{thebibliography}

\end{document}